\documentclass[12pt]{article}
\begin{document}
\title{The Prediction and Evidence for a New Particle - antiparticle Force
and Intermediary Particle}
\author{B.G. Sidharth\\
International Institute for Applicable Mathematics \& Information Sciences\\
Hyderabad (India) \& Udine (Italy)\\
B.M. Birla Science Centre, Adarsh Nagar, Hyderabad - 500 063
(India)}
\date{}
\maketitle
\begin{abstract}
We review some recent studies of particle-antiparticle behaviour
(including proton-antiproton) at high energies, which suggest a new
short lived interaction between them. We then examine the latest
evidence from the CDF team at Fermi Lab which bears out this result
to a little above $3 \sigma$ level.
\end{abstract}
\section{Introduction}
The supposedly unphysical negative energy solutions appear in
relativistic Quantum Mechanics, whether it be in the Klein-Gordon or
the Dirac equation. For the Klein-Gordon equation it was immediately
recognized that the second time derivative gives an extra degree of
freedom in the form of the negative energy solutions. These extra
solutions were interpreted by Pauli and Weisskopf in a Quantum Field
theoretical sense \cite{pauli}. There was an approach by Feshbach
and Villars \cite{feshbach} in terms of a two component Klein-Gordon
equation, in which they retained a particle sense. In this case it
turned out that the negative solutions would represent anti
particles.\\
Dirac tried to circumvent these difficulties by invoking an equation
that was first order in the time derivative, although at the expense
of introducing ultimately four component wave functions
\cite{dirac}. In spite of this, the negative energy electron problem
remained and Dirac had to introduce the negative energy sea which
was ostensibly filled up, while a positive energy electron could not
transit into the sea by virtue of the Pauli Exclusion Principle. In
this case an empty negative energy electron state or hole would
appear as a Positron. All this is well known.
\section{Negative Energy States}
We first make a few comments: Only the positive energy states alone
do not form a complete set for the Klein-Gordon or Dirac equations.
For localization a particle would have to be represented by a wave
packet consisting of both positive energy and negative energy
solutions. The minimum extension of such a localized packet is of
the order of the Compton wavelength (Cf.ref.\cite{feshbach}). This
means that at energies which are high enough that we are near the
Compton wavelength, we begin to encounter negative energies.\\
It must be emphasized that the Dirac electron itself has the
velocity $c$! Dirac himself explained that this is the case if we
carry over in a straightforward manner the concept of spacetime
points \cite{dirac}. He emphasized that our physical measurements
are averages over spacetime intervals. In fact Wigner and Salecker
have argued that usual spacetime points are invalid within the
Compton scale \cite{wigner}. Within the Compton scale we encounter
unphysical phenomena like negative energy solutions and
Zitterbewegung \cite{bgszbe}.\\
In the case of the Dirac equation, we can form meaningful
probability currents and sub luminal velocities using positive
energy solutions alone as long as we are outside the Compton
wavelength \cite{bd}. However as we approach the Compton scale we
begin to encounter negative energy solutions, or in the Feshbach-
Villars description, anti particles.\\
To proceed, let us write the Dirac wave function as
\begin{equation}
\psi = \left(\begin{array}{ll} \phi \\
\chi\end{array}\right),\label{3.27}
\end{equation}
where $\phi$ and $\chi$ are each two spinors. As is well known
\cite{bd}, we can then deduce
$$\imath \hbar (\partial \phi / \partial t) = c \tau \cdot (p -
e/cA) \chi + (mc^2+e\phi)\phi ,$$
\begin{equation}
\imath \hbar (\partial \chi / \partial t) = c \tau \cdot (p - e/cA)
\phi + (-mc^2+e\phi)\chi.\label{3.29}
\end{equation}
We recapitulate that at low energies $\chi$ is small and $\phi$
dominates, whereas it is the reverse at high energies. We also note
that while sensible wave packets can be formed with the positive
energy solutions alone, in general we require both signs of energy
for a localized particle. In fact the Compton wavelength is the
minimum extension, below which both positive and negative solutions
will have to be considered. Well outside the Compton wavelennth, we
can continue with the usual positive energy description. More
formally the positive energy solutions alone do not form a complete
set of eigen functions of the Hamiltonion.\\
The following symmetry can be seen from (\ref{3.29}) (with $e = 0$,
or the absence of an external electromagnetic field for simplicity):
\begin{equation}
t \to -t, \phi \to - \chi\label{3}
\end{equation}
We must remember that we are now dealing with intervals at the
Compton scale, so that the negative energy solutions are relevant.
So the
time reversal given in (\ref{3}) is at the Compton scale.\\
It has been pointed out by the author that this flip flop in time in
the microscopic interval $(t \to -t)$ can be modelled by a Double
Weiner process \cite{tduniv,ness}. It is related to Zitterbewegung (Cf.ref.\cite{bgszbe}).\\
It is quite remarkable that in the case of the Klein-Gordon equation
too, we get back equations identical to (\ref{3.29}), but this time
the wave functions have slightly different meaning. To see this in a
little greater detail, we look for an appropriate $\Psi$ such that
$\Psi$ satisfies the Hamiltonian form for the wave equation:
\begin{equation}
H\Psi = \imath \hbar (\partial \Psi / \partial t).\label{2.10}
\end{equation}
To obtain this form it is necessary to resolve $\Psi$ into the
components representing the two degrees of freedom implied by the
Klein-Gordon equation: $\phi$ and $\chi$. The function $\Psi$ is
then a unicolumn matrix formed from these
two components identical to (\ref{3.27}).\\
The obvious first step in such a development is to introduce
$\partial \psi / \partial t$ as an independent component. Let
(Cf.ref.\cite{feshbach})
\begin{equation}
\psi_4 = - k^{-1} D_4 \psi.\label{2.11}
\end{equation}
Then the Klein-Gordon equation may be written in an equivalent form
equivalent to as follows:
$$D_4 \psi + k\psi_4 = 0,$$
\begin{equation}
\sum_k D_k^2 \psi - k D_4 \psi_4 - k^2 \psi = 0.\label{2.12}
\end{equation}
where $D_\mu$ is the derivative with respect to the $\mu$th
coordinate. These equations are already in the Hamiltonian form
(\ref{2.10}), but the combination $\psi$ and $\psi_4$ does not prove
to be convenient because of the asymmetry of (\ref{2.12}).
Accordingly, we introduce the linear combination,
$$\psi = 1/\sqrt{2} (\phi + \chi),$$
\begin{equation}
\psi_4 = 1 \sqrt{2} (\phi - \chi).\label{2.13}
\end{equation}
The equations for $\phi$ and $\chi$ are
$$D_4 \phi = (1/2k) \sum_k D_k^2 (\phi + \chi) - k \phi,$$
\begin{equation}
D_4 \chi = (1/2k) \sum_k D_k^2 (\phi + \chi) + k \chi\label{2.14}
\end{equation}
which may be written more explicitly as
$$\imath \hbar (\partial \phi / \partial t) = (1/2m) (\hbar / \imath
\nabla - eA/c)^2 (\phi + \chi)$$
$$\quad \quad \quad \quad +(e \phi + mc^2 \phi$$
$$\imath \hbar (\partial \chi / \partial t) = - (1/2m) (\hbar / \imath
\nabla - eA/c)^2 (\phi + \chi)$$
\begin{equation}
\quad \quad \quad \quad + (e \phi - mc^2)\chi\label{2.15}
\end{equation}
very similar to (\ref{3.29}) of the Dirac case, except that in this
case $\phi$ and $\chi$ are scalar functions, rather than 2 spinors
as in the Dirac case.
\section{A New Interaction}
To proceed further we observe the the two Weiner process referred to
above, that is $(t \to -t)$ can be described by the following
equation
\begin{equation}
\frac{d_+}{dt} x (t) = {\bf b_+} \, , \, \frac{d_-}{dt} x(t) = {\bf
b_-}\label{2ex1}
\end{equation}
where we are for the moment in the one dimensional case. This
equation (\ref{2ex1}) expresses the fact that the right derivative
with respect to time is not necessarily equal to the left
derivative. It is well known that (\ref{2ex1}) leads to the Fokker
Planck equations \cite{tduniv,ness}
$$
\partial \rho / \partial t + div (\rho {\bf b_+}) = V \Delta \rho
,$$
\begin{equation}
\partial \rho / \partial t + div (\rho {\bf b_-}) = - U \Delta
\rho\label{2ex2}
\end{equation}
defining
\begin{equation}
V = \frac{{\bf b_+ + b_-}}{2} \quad ; U = \frac{{\bf b_+ - b_-}}{2}
\label{2ex3}
\end{equation}
We get on addition and subtraction of the equations in (\ref{2ex2})
the equations
\begin{equation}
\partial \rho / \partial t + div (\rho V) = 0\label{2ex4}
\end{equation}
\begin{equation}
U = \nu \nabla ln\rho\label{2ex5}
\end{equation}
It must be mentioned that $V$ and $U$ are the statistical averages
of the respective velocities and their differences. We can then
introduce the definitions
\begin{equation}
V = 2 \nu \nabla S\label{2ex6}
\end{equation}
\begin{equation}
V - \imath U = -2 \imath \nu \nabla (l n \psi)\label{2ex7}
\end{equation}
We will not pursue this line of thought here but refer the reader to
Smolin \cite{smolin} for further details. We now observe that the
complex velocity in (\ref{2ex7}) can be described in terms of a
positive or uni directional time $t$ only, but a complex coordinate
\begin{equation}
x \to x + \imath x'\label{2De9d}
\end{equation}
To see this let us rewrite (\ref{2ex3}) as
\begin{equation}
\frac{dX_r}{dt} = V, \quad \frac{dX_\imath}{dt} = U,\label{2De10d}
\end{equation}
where we have introduced a complex coordinate $X$ with real and
imaginary parts $X_r$ and $X_\imath$, while at the same time using
derivatives with respect
to time as in conventional theory.\\
From (\ref{2ex3}) and (\ref{2De10d}) it follows that
\begin{equation}
W = \frac{d}{dt} (X_r - \imath X_\imath )\label{2De11d}
\end{equation}
This shows that we can use derivatives with respect to the usual
time derivative with the complex space coordinates (\ref{2De9d}) (Cf.ref.\cite{bgsfpl162003}.\\
Generalizing (\ref{2De9d}), to three dimensions, we end up with not
three but four dimensions,
$$(1, \imath) \to (I, \tau),$$
where $I$ is the unit $2 \times 2$ matrix and $\tau$s are the Pauli
matrices. We get the special relativistic \index{Lorentz}Lorentz
invariant metric at the same time.\\
That is,\\
\begin{equation}
x + \imath y \to Ix_1 + \imath x_2 + jx_3 + kx_4,\label{Aa}
\end{equation}
where $(\imath ,j,k)$ momentarily represent the \index{Pauli}Pauli
matrices; and, further,
\begin{equation}
x^2_1 + x^2_2 + x^2_3 - x^2_4\label{B}
\end{equation}
is invariant.\\
The right hand side of (\ref{Aa}) in terms of \index{Pauli}Pauli
matrices, obeys the quaternionic algebra of the second rank
\index{spin}spinors (Cf.Ref.\cite{bgsfpl162003,shirokov,sachsgr} for details).\\
To put it briefly, the quarternion number field obeys the group
property and this leads to a number system of quadruplets as a
minimum extension.\\
In fact one representation of the two dimensional form of the
\index{quarternion}quarternion basis elements is the set of
\index{Pauli}Pauli matrices above. Thus a
\index{quarternion}quarternion may be expressed in the form
$$Q = -\imath \tau_\mu x^\mu = \tau_0x^4 - \imath \tau_1 x^1 - \imath \tau_2 x^2 -
\imath \tau_3 x^3 = (\tau_0 x^4 + \imath \vec \tau \cdot \vec r)$$
This can also be written as
$$Q = -\imath \left(\begin{array}{ll}
\imath x^4 + x^3 \quad x^1-\imath x^2\\
x^1 + \imath x^2 \quad \imath x^4 - x^3
\end{array}\right).$$
As can be seen from the above, there is a one to one correspondence
between a \index{Minkowski}Minkowski four-vector and $Q$. The
invariant is now given by
$Q\bar Q$, where $\bar Q$ is the complex conjugate of $Q$.\\
In this description we would have from (\ref{Aa}), returning to the
usual notation,
\begin{equation}
[x^\imath \tau^\imath , x^j \tau^j] \propto \epsilon_{\imath jk}
\tau^k \ne 0\label{y}
\end{equation}
(No summation over $\imath$ or $j$)\\
Equation (\ref{y}) shows that the coordinates no longer follow a
commutative geometry. It is quite remarkable that the noncommutative
geometry (\ref{y}) has been studied by the author in some detail
(Cf.\cite{tduniv}), though from the point of view of Snyder's
minimum fundamental length, which he introduced to overcome
divergence difficulties in Quantum Field Theory. Indeed we are
essentially in the same situation, because as we have seen, for our
positive energy description of the universe, there is the minimum
Compton wavelength cut off for a meaningful description
\cite{bgsextn,schweber,newtonwigner}.\\
Proceeding further we could think along the lines of $SU (2)$ and
consider the gauge transformation \cite{taylor}
\begin{equation}
\psi (x) \to exp [\frac{1}{2} \imath g \tau \cdot \omega (x)] \psi
(x).\label{4.2}
\end{equation}
This leads as is well known to a gauge covariant derivative
\begin{equation}
D_\lambda \equiv \partial_\lambda - \frac{1}{2} \imath g \tau \cdot
\bar{W}_\lambda,\label{4.3a}
\end{equation}
where $\omega$ in this theory is infinitessimal. We are thus lead to
vector Bosons $\bar{W}_\lambda$ and an interaction like the strong
interaction, described by
\begin{equation}
\bar{W}_\lambda \to \bar{W}_\lambda + \partial_\lambda \omega - g
\omega \Lambda \bar{W}_\lambda.\label{4.4}
\end{equation}
However, we are this time dealing, not with iso spin, but between
positive and negative energy states as in (\ref{3.27}) that is
particles and antiparticles. Also we must bear in mind that this
non-electromagnetic force between particles and anti particles would
be short lived as we are at the Compton scale \cite{report}.\\
These considerations are also valid for the Klein-Gordon equation in
the two component notation developed by Feshbach and Villars
\cite{feshbach,uheb}. There too, we get equations like (\ref{3.29}).
We would like to re-emphasize that our usual description in terms of
positive energy solutions is valid above the Compton scale.\\
Thus we are lead to a new short lived interaction (as we are near
the Compton scale), mediated by vector Bosons $\bar{W}$.\\
With regard to the $\bar{W}$ acquiring mass, apart from the usual
approach, we can note the following. Equation (\ref{y}) underlines
the non-commutativity of spacetime, and under these circumstances it
has been argued that there is a break in symmetry that leads to a
mass being acquired exactly as with the Higgs mechanism
\cite{bgsijmpe,tduniv}.
\section{Experimental Observation}
It is quite remarkable that evidence for such a reaction has just
been announced by the CDF team of Fermi Lab \cite{arxiv}. They
report a study of the invariant mass distribution of jet pairs
produced in association with a $W$ boson using data collected with
the CDF detector which correspond to an integrated luminosity of
$4.3 fb^{-1}$. The observed distribution has an excess in the
$120-160 GeV/c^2$ mass range which is not described by current
theoretical predictions within the statistical and systematic
uncertainties. In this latter they report studies of the properties
of this excess.\\
In fact what the authors, T. Aaltonen et al., have done is, they
performed a statistical comparison of the measurements of associated
production of $\bar{W}$ Boson and jets by including additional data
and further studying $M_{jj}$ distribution for masses higher than
$100 GeV$, with minimal changes to the event selection with respect
to the previous analysis. They found a statistically significant
disagreement that the other theoretical predictions.\\
Their model describes the data within uncertainties, except in the
mass region $\sim 120-160 GeV$ where an excess over the simulation
is seen. Briefly they have found evidence for a new Force and
Particle, unrelated to known physics in the Proton $p -\bar{p}$
interactions. Their results are accurate to the above $3 \sigma$
level. Further analysis is required to push the confidence levels to
the $5 \sigma$ level.
\section{Remarks}
As noted in Section 3, the non-commutativity (\ref{y}) can generate
mass. Let us see this in greater detail. The Gauge Bosons would be \index{mass}massless
and hence the need for a \index{symmetry breaking}symmetry breaking, \index{mass}mass generating mechanism.\\
The well known remedy for the above situation has been to consider,
in analogy with \index{superconductivity}superconductivity theory,
an extra phase of a self coherent system (Cf.ref.\cite{moriyasu} for
a simple and elegant treatment and also refs. \cite{jacob} and
\cite{taylor}). Thus instead of the \index{gauge field}gauge field
$A_\mu$, we consider a new phase adjusted \index{gauge field}gauge
field after the \index{symmetry}symmetry is broken
\begin{equation}
\bar{W}_\mu = A_\mu - \frac{1}{q} \partial_\mu \phi\label{Eex4}
\end{equation}
The field $\bar{W}_\mu$ now generates the \index{mass}mass in a self
consistent manner via a Higgs mechanism. Infact the kinetic energy
term
\begin{equation}
\frac{1}{2} |D_\mu \phi |^2\quad ,\label{Eex5}
\end{equation}
where $D_\mu$ in (\ref{Eex5}) denotes the Gauge derivative, now
becomes
\begin{equation}
|D_\mu \phi_0 |^2 = q^2|\bar{W}_\mu |^2 |\phi_0 |^2 \, ,\label{Eex6}
\end{equation}
Equation (\ref{Eex6}) gives the \index{mass}mass in terms of the ground state $\phi_0$.\\
The whole point is as follows: The \index{symmetry breaking}symmetry breaking of the \index{gauge field}gauge
field manifests itself only at short length scales signifying the fact that the field is mediated by particles
with large \index{mass}mass. Further the internal \index{symmetry}symmetry space of the \index{gauge field}gauge
field is broken by an external constraint: the wave function has an intrinsic relative phase factor which is a
different function of spacetime coordinates compared to the phase change necessitated by the minimum coupling
requirement for a free particle with the gauge potential. This cannot be achieved for an ordinary point like
particle, but a new type of a physical system, like the self coherent system of \index{superconductivity}superconductivity
theory now interacts with the \index{gauge field}gauge field. The second or extra term in (\ref{Eex4}) is effectively an
external field, though (\ref{Eex6}) manifests itself only in a relatively small spatial interval. The $\phi$ of the
Higgs field in (\ref{Eex4}), in analogy with the phase function of  \index{Cooper pairs}Cooper pairs of
\index{superconductivity}superconductivity theory comes with a \index{Landau-Ginzburg}Landau-Ginzburg potential $V(\phi)$.\\
Let us now consider in the \index{gauge field}gauge field
transformation, an additional phase term, $f(x)$, this being a
scalar. In the usual theory such a term can always be gauged away in
the \index{U(1)}U(1) \index{electromagnetic}electromagnetic group.
However we now consider the new situation of a
\index{noncommutative}noncommutative geometry referred to above,
\begin{equation}
\left[dx^\mu , dx^\nu \right] = \Theta^{\mu \nu} \beta , \beta \sim
0 (l^2)\label{Eex7}
\end{equation}
where $l$ denotes the minimum \index{spacetime}spacetime cut off.
Equation (\ref{Eex7}) is infact \index{Lorentz}Lorentz covariant.
Then the $f$ phase factor gives a contribution to the second order
in coordinate differentials,
$$\frac{1}{2} \left[\partial_\mu B_\nu - \partial_\nu B_\mu \right] \left[dx^\mu , dx^\nu \right]$$
\begin{equation}
+ \frac{1}{2} \left[\partial_\mu B_\nu + \partial_\nu B_\mu \right]
\left[dx^\mu dx^\nu + dx^\nu dx^\mu \right]\label{Eex8}
\end{equation}
where $B_\mu \equiv \partial_\mu f$.\\
As can be seen from (\ref{Eex8}) and (\ref{Eex7}), the new
contribution is in the term which contains the commutator of the
coordinate differentials, and not in the symmetric second term.
Effectively, remembering that $B_\mu$ arises from the scalar phase
factor, and not from the non-Abelian \index{gauge field}gauge field,
$A_\mu$ is replaced by
\begin{equation}
A_\mu \to A_\mu + B_\mu = A_\mu + \partial_\mu f\label{Eex9}
\end{equation}
Comparing (\ref{Eex9}) with (\ref{Eex4}) we can immediately see that the effect of noncommutativity is
precisely that of providing a new \index{symmetry breaking}symmetry breaking term to the \index{gauge field}gauge
field, instead of the $\phi$ term, (Cf.refs. \cite{cr39,ijmpe}) a term not belonging to the \index{gauge field}gauge field itself.\\
On the other hand if we neglect in (\ref{Eex7}) terms $\sim l^2$,
then there is no extra contribution coming from (\ref{Eex8}) or
(\ref{Eex9}), so that we are in the usual non-Abelian \index{gauge
field}gauge field theory, requiring a broken
\index{symmetry}symmetry to obtain an equation like (\ref{Eex9}).
This is not surprising because if we neglect the term $\sim l^2$ in
(\ref{Eex7}) then we are back with the usual commutative theory and
the usual \index{Quantum Mechanics}Quantum Mechanics.


\begin{thebibliography}{99}
\bibitem {pauli} Pauli, W and Weisskopf, V.F.. (1934). \emph{Helv.Phys.Acta 1}, 709 (1934).
\bibitem {feshbach} Freshbach, H. and Villars, F. (1958).
\emph{Rev.Mod.Phys.} Vol.30, No.1, January 1958, pp.24-45.
\bibitem {dirac} Dirac, P.A.M. (1958). \emph{The Principles of
Quantum Mechanics} (Clarendon Press, Oxford), pp.4ff, pp.253ff.
\bibitem {wigner} Salecker, H. and Wigner, E.P. (1958). \emph{Quantum
Limitations of the Measurement of Space-Time Distances}
\emph{Physical Review} Vol.109, No.2, January 15 1958, pp.571-577.
\bibitem {bgszbe} Sidharth, B.G. (2008). \emph{Int.J.Th.Phys.} 48
(2), 2008, pp.497-506.
\bibitem {bd} Bjorken, J.D. and Drell, S.D. (1964). \emph{Relativistic Quantum Mechanics}
(Mc-Graw Hill, New York), pp.39.
\bibitem {tduniv} Sidharth, B.G. (2008). \emph{The Thermodynamic
Universe} (World Scientific), Singapore.
\bibitem {ness} Sidharth, B.G. (2011). \emph{Negative Energy
Solutions and Symmetries}, \emph{arXiv:1104.0116}.
\bibitem {smolin} Smolin, L. (1986). \emph{Quantum Concepts in Space and Time},
Penrose, R. and  Isham, C.J. (eds.) (OUP, Oxford), pp.147--181.
\bibitem {bgsfpl162003} Sidharth, B.G. (2003) \emph{Found.Phys.Lett.} 16, (1), pp.91--97.
\bibitem {shirokov} Shirokov, Yu. M. (1958). \emph{Soviet Physics JETP} \underline{6}, (33),
No.5, pp.929--935.
\bibitem {sachsgr} Sachs, M. (1982). \emph{General Relativity and Matter} (D. Reidel
Publishing Company, Holland), pp.45ff.
\bibitem {bgsextn} Sidharth, B.G. (2002). \emph{Foundation of Physics Letters} 15 (5), 2002, 501ff.
\bibitem {schweber} S.S. Schweber. (1961). \emph{An Introduction to Relativistic Quantum
Field Theory} (Harper and Row, New York).
\bibitem {newtonwigner} Newton, T.D. and Wigner, E.P. (1949). \emph{Reviews of Modern Physics} Vol.21,
No.3, July 1949, pp.400-405.
\bibitem {taylor} Taylor, J.C. (1976). \emph{Gauge Theories of Weak
Interactions} (Cambridge University Press, Cambridge) 1976.
\bibitem {report} Sidharth, B.G. (2011). \emph{Brief Report} in
\emph{New Advances in Physics} Vol.5 (1), 2011, pp.49-50.
\bibitem {uheb} Sidharth, B.G. (2011) \emph{Ultra High Energy
Behaviour}, \emph{arXiv:1103.1496}.
\bibitem {bgsijmpe} Sidharth, B.G. (2005). \emph{Int.J.Mod.Phys.E.}
14 (6), 2005, pp.923ff.
\bibitem {arxiv}Aaltonen, T. et al. \emph{arXiv:1104.0699v1} in
\emph{hep-ex} 4 April 2011.
\bibitem {moriyasu}  Moriyasu, K. (1983). \emph{An Elementary Primer for Gauge Theory} (World Scientific, Singapore, 1983).
\bibitem {jacob} Greiner, W., and Muller, B. (2009). \emph{Gauge Theory of Weak Interactions} (Springer, New York, 2009).
\bibitem {cr39}  Sidharth, B.G. (2004). \emph{Proceedings of the Fifth International Symposium} on \emph{Frontiers of Fundamental Physics} in
(Universities Press, Hyderabad, 2004).
\bibitem {ijmpe} Sidharth, B.G. (2005). \emph{Int.J.Mod.Phys.E}
14 (2), 2005, p.215ff.
\end{thebibliography}
\end{document}